\documentclass[10pt]{article}

\usepackage[english]{babel}

\usepackage{graphics,color}
\usepackage{newlfont,epsfig}

\begin{document}
\date{}
\begin{center}
{\Large\bf Generation of hybrid entanglement between a mirror and a trapped ion}
\end{center}
\begin{center}
{\normalsize Cl\'ovis Corr\^ea and A. Vidiella-Barranco \footnote{vidiella@ifi.unicamp.br}}
\end{center}
\begin{center}
{\normalsize{ Instituto de F\'\i sica ``Gleb Wataghin'' - Universidade Estadual de Campinas}}\\
{\normalsize{ 13083-859   Campinas  SP  Brazil}}\\
\end{center}
\begin{abstract}
We present a scheme for cavity-assisted generation of hybrid entanglement between a moving mirror belonging to an 
optomechanical cavity and a single trapped ion located inside a second cavity. Due to radiation pressure, it is 
possible to entangle the states of the moving mirror and the corresponding cavity field. Also, by tuning the second cavity 
field with the internal degrees of freedom of the ion, an entangled state of the cavity field/ion can be 
independently generated. The fields leaking from each cavity may be then combined in a beam-splitter, and following 
the detection of the outgoing photons by conveniently placed photodetectors, we show that it is possible to generate 
entangled states of the moving mirror and the single trapped ion's center-of-mass vibration. In our scheme the generated 
states are {\it hybrid entangled states}, in the sense that they are constituted by discrete (Fock) states and continuous 
variable (coherent) states.
\end{abstract}
\section{Introduction}

The two fast growing fields of cavity optomechanics \cite{marq14} and trapped ions \cite{wine03} have been opening new avenues
for the investigation of quantum phenomena. In particular, one may consider the confinement of a trapped ion in a cavity \cite{knight97} 
which would allow, for instance, the coupling of the quantized cavity field with the trapped 
ion itself \cite{blatt03,blatt12,blatt13}. This forms a bipartite quantum system constituted by a massive particle (ion)
and the electromagnetic field. Among the possible applications related to the trapped ion system, we could cite 
proposals for quantum information processing, e.g., quantum computing using laser driven trapped ions \cite{cirac95}, as well 
as cavity-assisted quantum computing \cite{avb99}. Needless to say that a high degree of control over the ionic dynamics has been
achieved, allowing the generation of nonclassical states \cite{wineland96a} as well as the reconstruction of the density 
matrix \cite{wineland96b}. Another interesting massive system is the one involving a ``macroscopic"  
object (mirror) interacting with the electromagnetic field. In fact, in such an optomechanical cavity, 
it is possible to couple the moving mirror (via radiation pressure) with the cavity field \cite{walls93,kippen12}. 
Considering the trapped ion-field and mirror-field systems independently, the involved interactions naturally lead, on one side, 
to the entanglement between the vibrating ion and the field (cavity $1$) \cite{knight97,avb01} and on the other side, the 
entanglement between the oscillating mirror and the field (cavity $2$) \cite{tombesi97,bose97}. 
We could thus think about the feasibility of generating entanglement between the (microscopic) ionic subsystem and the 
(macroscopic) mirror via the fields. Regarding the optomechanical cavities, we should point out that there are already 
proposals for the generation of entanglement in a system involving two cavities; more specifically, between two moving 
mirrors belonging to distinct cavities \cite{tombesi02,vedral08,nori14}. In particular, we highlight a scheme of
entanglement generation involving two optomechanical cavities in the realm of the single-photon regime \cite{nori14}. 

Here we would like to discuss the possibility of entangling the moving ion located within cavity $1$, with the moving mirror 
in cavity $2$, also in the single-photon regime \cite{nori14}, using an interferometry-based 
scheme of the outgoing fields similar to the one presented in \cite{vedral99}.  
We show that the generation of ion-mirror entanglement may be achieved under certain conditions, and that the 
resulting entangled states have component states involving (discrete) Fock and (continuous variable) coherent states, i.e., 
they are hybrid entangled states \cite{vanloock12}. The concept of hybrid entanglement is also relevant for 
quantum information processing purposes. For instance, we may find in the literature proposals of quantum gates \cite{wang01} 
as well as quantum repeaters \cite{vanloock06,long13} based on hybrid (discrete/continuous variable) states. 
We should mention the existence of previous (but different) works presenting proposals for generating entanglement between trapped ions 
and mechanical oscillators. For instance, schemes based on electrostatic coupling are presented in \cite{zoller04,milburn05}.
Other possibilities are also discussed in the review paper \cite{treutlein11}; we highlight a proposal involving the coupling of
an atom with a mechanical oscillator in a cavity \cite{kimble09}. However in those proposals the atom (or ion) and the mechanical 
oscillator must be very close, while in our light-mediated scheme the corresponding quantum systems may be located 
in cavities which are far apart. Note that an indirect coupling should also be suitable for entangling 
sub-systems having a considerable mass difference \cite{treutlein11}, e.g., a trapped ion and a mirror.
Our paper is organized as follows: in Section 2 we review the dynamics of the ion trapped in a cavity as well as the 
optomechanical cavity; in Section 3 we present our scheme for the ion-mirror entanglement, and in Section 4 we discuss the
amount of entanglement in the generated states. In Section 5 we summarize our conclusions.

\section{Models for the trapped ion and the optomechanical systems}

We consider, as shown in Fig. (1), an interferometer having two arms; a cavity containing a single (two-level) ion 
confined inside a Paul trap is positioned in arm 1 and an optomechanical cavity with a moving mirror is placed in arm 2.
Firstly we are going to describe the preparation of the ionic and optomechanical systems, 
which consists in the independent preparation of an entangled state of the ion with the field in cavity 1,
and the preparation of an entangled state involving the moving mirror with the field in cavity 2.

\begin{figure}
\centering
\resizebox{0.8\textwidth}{!}{
\includegraphics{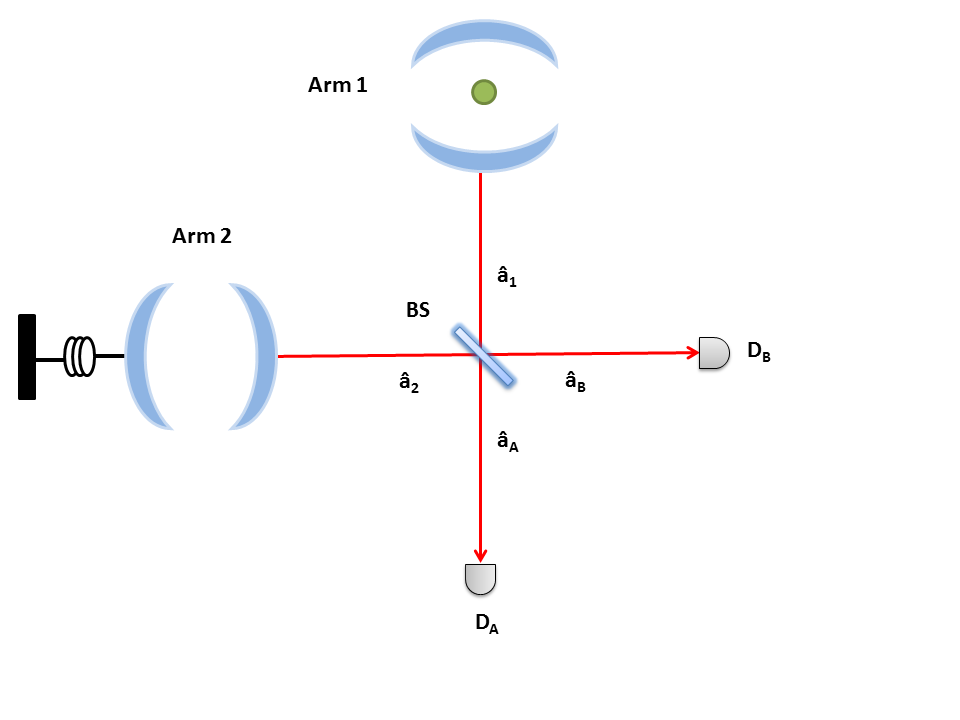}
}
\caption{Illustration of a cavity with a trapped ion (green circle) at arm 1, and an optomechanical cavity at arm 2. 
If the radiation fields from both arms are simultaneously focused at the $50:50$ beam splitter, an 
entangled state between the ionic vibrational state and the moving mirror is generated after the detection of a photon.}
\end{figure}


\subsection{Trapped ion in a cavity}

The Hamiltonian describing the dynamics of the quantized cavity field coupled to the ion (vibrational motion + internal electronic states)  
may be written as \cite{knight97,avb01}
\begin{equation}
\hat{H}_{ion}=\hbar\omega_{c}\hat{a}^{\dagger}\hat{a}+\hbar\omega_{v}\hat{b}^{\dagger}\hat{b}+\hbar\frac{\omega_{0}}{2}\hat{\sigma}_{z}
	+\hbar G(\hat{\sigma}_{+}+\hat{\sigma}_{-})(\hat{a}+\hat{a}^{\dagger})\sin\eta(\hat{b}+\hat{b}^{\dagger}),\label{hamilcomplete}
\end{equation}
where $\hat{a}$($\hat{a}^\dagger$) denotes the annihilation (creation) operator of the cavity field (with frequency $\omega_{c}$), 
$\hat{b}$($\hat{b}^\dagger$) is the annihilation (creation) operator of the vibrational degree of freedom of the center-of-mass of the 
ion (with frequency $\omega_{v}$), $\omega_{0}$ it is the atomic frequency, $G$ is the ion-field coupling constant, and 
$\eta = 2\pi a_0/\lambda$ is the Lamb-Dicke parameter; $a_0$ being the characteristic amplitude of the ion's harmonic motion 
and $\lambda$ is the wavelength of the cavity field. The Hamiltonian above may be linearized if the ion is confined in a region typically 
much smaller than the wavelength of light, or $\eta\ll 1$ (Lamb-Dicke regime).

\subsubsection{Trapped ion with quantum jumps: first red sideband case}

An adequate way of treating the photon leaking process (from the cavities) is via the quantum jump formalism 
\cite{vedral99,plenio98,huelga03,zheng08}. The effective Hamiltonian of the cavity (having decay rate $\gamma$) with a trapped ion 
governing the evolution while there is no photon detection, is given by
\begin{equation}
	\hat{\cal{H}}_{ion}=\hat{H}_{ion}-i\frac{\hbar\gamma}{2}\hat{a}^{\dagger}\hat{a}.\label{jumpion1}
\end{equation}

If we tune the system to the first red sideband ($\delta=\omega_{0}-\omega_{c}=\omega_{v}$), we obtain, in the interaction picture, 
\begin{equation}
	\hat{\cal{H}}_{ion,I}^{r}=\eta\hbar G(\hat{\sigma}_{-}\hat{a}^{\dagger}\hat{b}^{\dagger}+\hat{\sigma}_{+}\hat{a} \hat{b})-i\frac{\hbar\gamma}{2}\hat{a}^{\dagger}\hat{a}.
\end{equation}

Thus the associated non-Hermitian evolution operator is
\begin{equation}
	{\cal{U}}_{ion}^{r}(t')={\exp}\left[-i\frac{\hat{\cal{H}}_{ion,I}^{r}}{\hbar}t'\right].\label{evopred}
\end{equation}

We may choose the initial joint ion-cavity state as being
\begin{equation}
	|\psi_{ion}(0)\rangle=|0\rangle_{1}|0\rangle_{v}(\cos\theta|e\rangle+\sin\theta|g\rangle),\label{initialioncavity}
\end{equation}
where $|0\rangle_{1}$ e $|0\rangle_{v}$ are the cavity field state (arm 1) and the vibrational ionic state, respectively, 
prepared in their vacuum states. We remark that in order to generate an entangled (ion + field) state,  the ion's electronic states 
must be prepared in a superposition of the ``ground" state $|g\rangle$ and the ``excited" state $|e\rangle$ \cite{avb01},
which could be done by applying convenient $\pi/2$ Raman pulses to the ion \cite{wine03}. Nevertheless,
it is required a simple preparation of both the ionic motion \cite{wine03} and the cavity; they just need to be cooled down to their 
ground states. 

The evolution of initial state in Eq.(\ref{initialioncavity}) under (\ref{evopred}) followed by a suitable measurement of the ion's 
electronic state results in (see Appendix A)

\begin{equation}
|\psi_{ion,g}^{r}(t'=\pi/2\eta G)\rangle= 
N^{(r)} \Bigl[\sin\theta|0\rangle_{1}|0\rangle_{v}-i\cosh\left(\frac{\pi^{2}\gamma}{8\eta G}\right)\cos\theta|1\rangle_{1}|1\rangle_{v}\Bigr],
\label{ionfieldstate}
\end{equation}
where $N^{(r)}$ is a normalizing constant.
We may calculate the probability that there is no photodetection during the evolution, which alows the generation of the state in 
Eq.(\ref{ionfieldstate}), or 
\begin{eqnarray}
	P_{\mbox{ND}}^{r}(ion)&=&|\langle g|\psi_{ion}^{r}(t'=\pi/2\eta G)\rangle|^{2} \nonumber \\
	&=&\left[1+\frac{e^{-\frac{\pi\gamma}{\eta G}}\sinh^{2}\left(\frac{\pi^{2}\gamma}{8\eta G}\right)}{\cosh^{2}
	\left(\frac{\pi^{2}\gamma}{8\eta G}\right)+\tan^{2}\theta}\right]^{-1}.
\end{eqnarray}

\subsubsection{Trapped ion with quantum jumps: first blue sideband case}

The ion-field detuning may also be adjusted to the first blue sideband, ($\delta=\omega_{0}-\omega_{c}=-\omega_{v}$), 
leading to the following interaction picture Hamiltonian (with quantum jump),

\begin{equation}
\hat{\cal{H}}_{ion,I}^{b}=\eta\hbar G(\hat{\sigma}_{-}\hat{a}\hat{b}^{\dagger}+\hat{\sigma}_{+}\hat{a}^{\dagger} 
\hat{b})-i\frac{\hbar\gamma}{2}\hat{a}^{\dagger}\hat{a},
\end{equation}

having the corresponding evolution operator
\begin{equation}
	{\cal{U}}_{ion}^{b}(t')={\exp}\left[-i\frac{\hat{\cal{H}}_{ion,I}^{b}}{\hbar}t'\right].
\end{equation}

The cavity should be initially cooled to its vacuum state $|0\rangle_1$, but contrarily to the previous case (first red sideband), the ion 
should be initially prepared in its first excited state, or $|1\rangle_{v}$ \cite{wine03}. Thus

\begin{equation}
	|\psi_{{ion}}^{{b}}(t')\rangle=\,{\cal{U}}_{{ion}}^{{b}}(t')|0\rangle_{1}|1\rangle_{v}(\cos\theta|e\rangle
	+\sin\theta|g\rangle).
\end{equation}

Following the same steps as in the red sideband case, we obtain (see Appendix A)

\begin{equation}
|\psi_{ion,g}^{b}\left(t'=\pi/2\eta G\right)\rangle=N^{(b)}\left[\sin\theta|0\rangle_{1}|1\rangle_{v}-ie^{-\frac{\gamma t}{2}}
\cosh\left(\frac{\pi^{2}\gamma}{8\eta G}\right)\cos\theta|1\rangle_{1}|0\rangle_{v}\right]\label{cavityionblue}
\end{equation}
being $N^{(b)}$ a normalizing constant.

\subsection{Optomechanical cavity}

Now we turn our attention to arm 2, which contains the optomechanical cavity with a moving mirror. The Hamiltonian for such a 
system is given by \cite{bose97}
\begin{equation}
	\hat{H}_{om}=\hbar\Omega_{c}\hat{A}^{\dagger}\hat{A}+\hbar\Omega_{m}\hat{B}^{\dagger}\hat{B}-
	\hbar g\hat{A}^{\dagger}\hat{A}(\hat{B}+\hat{B}^{\dagger}).\label{hamilopto}
\end{equation}
Here $\hat{A}$, $\hat{B}$ are the annihilation operator of the cavity field and the moving mirror 
(modeled as a simple harmonic oscillator), respectively, and $g$ is the moving mirror-cavity field coupling constant,  
$g=\frac{\Omega_{c}}{L}\sqrt{\frac{\hbar}{2m\Omega_{m}}}$, being $L$ the length of the cavity and $m$ the mirror's mass.

\subsubsection{Optomechanical cavity with quantum jumps}

For an optomechanical cavity (decay rate $\Gamma$), the effective Hamiltonian in the quantum jump formalism will read

\begin{equation}
	\hat{{\cal{H}}}_{om}=\hat{H}_{om}-i\frac{\hbar\Gamma}{2}\hat{A}^{\dagger}\hat{A},
\end{equation}
where $\hat{H}_{om}$ is the optomechanical Hamiltonian in Eq.(\ref{hamilopto}). Thus,
\begin{equation}
	\hat{{\cal{U}}}_{om}(t'')={\exp}\left[-i\frac{\hat{{\cal{H}}}_{om}}{\hbar}t''\right].
\end{equation}

We assume the following specific initial state for the optomechanical system
\begin{equation}
	|\psi_{om}(0)\rangle=\frac{1}{\sqrt{2}}(|0\rangle_{2}+|1\rangle_{2})|\alpha_{0}\rangle_{m},\label{initialopto}
\end{equation}
where $|\varphi(0)\rangle = |0\rangle_{2}+|1\rangle_{2}$ and $|\alpha_{0}\rangle_{m}$ 
are the initial cavity field (superposition) state and the initial (coherent) state of the  
moving mirror, respectively. Such an initial state could be prepared in the following way; 
the mirror should be first cooled down to its ground state \cite{simmonds11,painter11}. Then a coherent state of
the oscillating mirror may be generated via the application of a light pulse (displacement) \cite{marq14,aspelmeyer11}.
The field of the corresponding cavity has to be prepared in a superposition of the vacuum state and the one-photon state, or 
$|\Phi\rangle \propto |0\rangle + |1\rangle$. We find in the literature various proposals of generation of such cavity superposition 
states, for instance, using travelling atoms, as discussed in \cite{haroche98}. 

Thus the evolved optomechanical state vector will read
\begin{eqnarray}
	|\psi_{om}(t'')\rangle&=&\hat{{\cal{U}}}_{om}(t'')|\psi_{om}(0)\rangle \nonumber \\
	&=&M \Bigl[|0\rangle_{2}|\alpha_{0}e^{-i\Omega_{m}t''}\rangle_{m} +
	e^{-\frac{\Gamma}{2}t''}e^{i\phi(t'')}|1\rangle_{2}|\alpha(t'')\rangle_{m}\Bigr],
\end{eqnarray}
with $\phi(t'')=\kappa^{2}(\Omega_{m}t''-\sin\Omega_{m} t'')$, $\alpha(t'')=\alpha_{0}e^{-i\Omega_{m}t''}+\kappa(1-e^{-i\Omega_{m}t''})$, 
$\kappa=g/\Omega_{m}$, and $M$ is a normalizing constant.

Now for $t''=\pi/\Omega_{m}$, we obtain
	\begin{equation}
	|\psi_{om}(t''=\pi/\Omega_{m})\rangle = M\Bigl[|0\rangle_{2}|-\alpha_{0}\rangle_{m} +
	e^{-\frac{\pi\Gamma}{2\Omega_{m}}}e^{i\pi\kappa^{2}}|1\rangle_{2}|-\alpha_{0}+2\kappa\rangle_{m}\Bigr],\label{mirrorfieldstate}
\end{equation}
which corresponds to the preparation of the optomechanical cavity prior to the detection stage.

\section{Generation of hybrid entanglement between the ionic and optomechanical systems}

In order to establish entanglement between the ion and the moving mirror subsystems, the fields from each cavity 
are combined in a beam-splitter \cite{vedral99,huelga03}, as shown in Fig.(1). After impinging the beam splitter,
photons may be detected by the photocounters $D_A$ or $D_B$.

\subsection{Ion-mirror entanglement: first red sideband case}

Firstly we are going to consider the first red sideband preparation for the ionic system.
If no photons are detected by $D_A$ (or $D_B$), the joint state in the beginning of the detection stage 
will be the product of state $|\psi_{ion,g}^{r}(t'=\pi/2\eta G)\rangle$, the ionic-cavity field state (arm 1) in 
Eq.(\ref{ionfieldstate}) and state $|\psi_{om}(t''=\pi/\Omega_{m})\rangle$, the mirror-cavity field (arm 2) in 
Eq.(\ref{mirrorfieldstate}). Naturally it is required a synchronization of the actions during the process.
We assume a finite detection time $t_{D}$, and thus for $t\leq t_{D}$ the system's joint state may be written as

\begin{equation}
	|\chi(t)\rangle=|\psi_{ion,g}^{r}(t)\rangle\otimes|\psi_{om}(t)\rangle, 
\end{equation}
where
\begin{eqnarray}
|\psi_{ion,g}^{r}(t)\rangle&=& {\exp}\left[-\frac{\gamma}{2}\hat{a}^{\dagger}\hat{a}t\right]|\psi_{{ion,g}^{r}}(t'=\pi/2\eta G)\rangle \nonumber \\
&=& \sin\theta|0\rangle_{1}|0\rangle_{v}
-ie^{-\frac{\gamma}{2}t}\cosh\left(\frac{\pi^{2}\gamma}{8\eta G}\right)\cos\theta|1\rangle_{1}|1\rangle_{v},
\end{eqnarray}
and
\begin{eqnarray}
|\psi_{{om}}(t)\rangle&=& \, {\exp}\left[-\frac{\Gamma}{2}\hat{A}^{\dagger}\hat{A}t\right]|\psi_{om}(t''=\pi/\Omega_{m})\rangle \\
&=& \frac{1}{\sqrt{2}}|0\rangle_{2}|-\alpha_{0}\rangle_{m}+\frac{1}{\sqrt{2}}e^{-\frac{\Gamma}{2}t}e^{-\frac{\pi\Gamma}{2\Omega_{m}}}
e^{i\pi\kappa^{2}}|1\rangle_{2}|-\alpha_{0}+2\kappa\rangle_{m}.\nonumber
\end{eqnarray}

After the fields from each cavity cross the beam-splitter, if a photon is detected by $D_B$, (field state projected onto 
$|0\rangle_{A}|1\rangle_{B}$), we obtain
\begin{equation}
|\Psi^r_{01}(t)\rangle = N^{(r)}_{01}\biggl[C^{(r)}_1(t) |0\rangle_{v}|-\alpha_{0}+2\kappa\rangle_{m} 
+ C^{(r)}_2(t)|1\rangle_{v}|-\alpha_{0}\rangle_{m} \biggr],\label{entangstred}
\end{equation}
with 
\begin{equation}
C^{(r)}_1(t) = e^{-\frac{\Gamma t}{2}}e^{-\frac{\pi\Gamma}{2\Omega_{m}}}e^{i\pi\kappa^{2}},
\end{equation}
and 
\begin{equation}
C^{(r)}_2(t) = e^{-\frac{\gamma t}{2}}\cosh\left(\frac{\pi^{2}\gamma}{8\eta G}\right)\cos\theta.
\end{equation}

Now if a photon is detected by $D_A$ (field state projected onto $|1\rangle_{A}|0\rangle_{B}$), the resulting state is
\begin{equation}
	|\Psi^r_{10}(t)\rangle = N^{(r)}_{10}\biggl[C^{(r)}_1(t) |0\rangle_{v}|-\alpha_{0}+2\kappa\rangle_{m} - 
	C^{(r)}_2(t)|1\rangle_{v}|-\alpha_{0}\rangle_{m} \biggr] .
\end{equation} 

We have therefore demonstrated the possibility of generation of hybrid entangled states (Fock/coherent) in subsystems 
belonging to spatially separated cavities.
  
\subsection{Ion-mirror entanglement: first blue sideband case}

In what follows we discuss the first blue sideband case for the ionic system, with
the optomechanical cavity having the same preparation as in the previous case [see Eq. (\ref{mirrorfieldstate})].
Analogously to the red sideband case, we obtain from Eq. (\ref{cavityionblue})

\begin{eqnarray}
|\psi_{ion,g}^{b}(t)\rangle&=&{\exp}\left[-\frac{\gamma}{2}\hat{a}^{\dagger}\hat{a}t\right]|
\psi_{ion,g}^{b}(t'=\pi/2\eta G)\rangle \nonumber \\
&=&\Bigl[e^{i\xi}\sin\theta|0\rangle_{1}|1\rangle_{v}
-ie^{-\frac{\gamma t}{2}}\cosh\left(\frac{\pi^{2}\gamma}{8\eta G}\right)\cos\theta|1\rangle_{1}|0\rangle_{v}\Bigr]
\end{eqnarray}
After the fields cross the beam-splitter, the subsequent detection of a photon by $D_B$ will result in the following state
\begin{equation}
|\Psi^b_{01}(t)\rangle = N^{(b)}_{01}\biggl[C^{(b)}_1(t) |0\rangle_{v}|-\alpha_{0}\rangle_{m} 
+ C^{(b)}_2(t) |1\rangle_{v}|-\alpha_{0}+2\kappa\rangle_{m}\biggr],
\end{equation}
where 
\begin{equation}
C_{1}^{(b)}(t)=\frac{1}{2}e^{-\frac{\gamma t}{2}}\cosh\left(\frac{\pi^{2}\gamma}{8\eta G}\right)\cos\theta
\end{equation}
and
\begin{equation}
C_{2}^{(b)}(t)=\frac{1}{2}e^{-\frac{\Gamma t}{2}}e^{-\frac{\pi\Gamma}{2\Omega}}e^{i\pi\kappa^{2}}\sin\theta
\end{equation}

If a photon is detected by $D_A$, we obtain

\begin{equation}
|\Psi^b_{10}(t)\rangle = N^{(b)}_{10}\biggl[C^{(b)}_1(t) |0\rangle_{v}|-\alpha_{0}\rangle_{m} 
- C^{(b)}_2(t) |1\rangle_{v}|-\alpha_{0}+2\kappa\rangle_{m}\biggr].
\end{equation}

Again, we have shown that entangled states of the hybrid type may be generated also for a different choice of frequency detunings.

\section{Degree of Entanglement}

An interesting peculiarity of the entangled states above is that, even if the mirror is initially just cooled down to its 
ground vibrational state ($\alpha_0 = 0$), the resulting states will be entangled. For instance, 
\begin{equation}
|\phi\rangle = A|0\rangle|2\kappa\rangle \pm B |1\rangle|0\rangle 
\end{equation} 
for the red sideband case, and
\begin{equation}
|\varphi\rangle = A'|0\rangle|0\rangle \pm B' |1\rangle|2\kappa\rangle
\end{equation} 
for the blue sideband case. 
We also note that if $\kappa = \alpha_0$, the generated states will be of the type
\begin{equation}
|\chi\rangle = A''|0\rangle|\pm\alpha_0\rangle \pm B'' |1\rangle|\mp\alpha_0\rangle 
\end{equation} 
which are entangled states relevant for quantum computation purposes \cite{milburn06,vanloock11,bellini14}.
Now we would like to discuss the degree of entanglement, for instance, of the state
generated in the red sideband case [Eq. (\ref{entangstred})]. Firstly we define an orthogonal basis for the continuous variable 
sub-space (via Gram-Schmidt process) as
\begin{equation}
	|\psi_{a}\rangle_{m}=|-\alpha_{0}\rangle_{m}
\end{equation}
and
\begin{equation}
	|\psi_{b}\rangle_{m}=\frac{1}{\sqrt{1-e^{-4\kappa^{2}}}}\left[|-\alpha_{0}+2\kappa\rangle_{m}-e^{\varphi}|-\alpha_{0}\rangle_{m}\right]
\end{equation}
Where $\varphi=\frac{1}{2}\left(|\alpha_0|^{2}-4\kappa^{2}-2\kappa(\alpha_0^{\ast}-\alpha_0)\right)$.
We may rewrite the state in Eq. (\ref{entangstred}) in terms of the states above as
\begin{equation}
	|\Psi_{01}^{r}(t)\rangle=N_{01}^{(r)}\left[C_{1}^{(r)}(t)\sqrt{1-e^{-4\kappa^{2}}}|0\rangle_{v}|\psi_{b}\rangle_{m}+C_{1}^{(r)}(t)e^{\varphi}|0\rangle_{v}|\psi_{a}\rangle_{m}+C_{2}^{(r)}(t)|1\rangle_{v}|\psi_{a}\rangle_{m}\right].\label{entangbasis}
\end{equation} 

The negativity of the state above may be calculated via the usual definition \cite{werner02}
\begin{equation}
	{\cal{N}}(\rho)=2\left|\sum_{i}\lambda_{i}^{-}\right|,
\end{equation}
being $\lambda_{i}^{-}$ the negative eigenvalues of the partial transpose of the density matrix associated to the state (\ref{entangbasis}).
In Fig. (2) we plot the negativity as a function of the coherent amplitude $\alpha_0$. In this specific case, for which $\kappa = 0.5$, it is 
clear that the state has maximum entanglement for $\alpha_0 = 0$. As expected, for larger values of $\alpha_0$, the ion-mirror state becomes 
separable.

\section{Discussion and conclusions}

We have presented a scheme for entanglement generation between a moving mirror and a trapped ion assisted by the quantized 
fields of two cavities. As a first step, the mirror as well as the ion should independently couple to the fields of their own cavities. 
After an adequate preparation of the initial states, the corresponding outgoing cavity fields are combined in a beam-splitter. Then a 
subsequent photodetection would collapse the global state, making possible the generation of entangled states of the ion and 
mirror subsystems. Note that in order to accomplish the entangled state generation, one should wait (finite time) for a single click in 
either detector A or detector B.

Concerning the time scales involved, despite of the fact that the quantum sub-systems we are dealing with are of a very different nature, 
namely, a ``microscopic" ion and a ``macroscopic" mirror, both of them could in principle have equal (or very close) oscillating 
frequencies, e.g., in the MHz range \cite{marq14,wine03,treutlein11}. We would like to remark that, due to the large mass difference 
between the ion and the mirror, a sort of ``impedance mismatch" occurs in schemes in which the ion is directly coupled to the mirror 
\cite{treutlein11}. Nevertheless, such a shortcoming is avoided in our light-mediated proposal, as the quantum sub-systems (ion/mirror) 
are indirectly coupled via the electromagnetic fields. 

Yet, there are sources of decoherence caused by dephasing in the ionic system, mechanical damping in the mirror as well as cavity losses. 
The mechanical damping (mirror) and the dephasing (ion) rates are much smaller than the cavity damping rate, $\gamma_{cav}$. 
Thus, in order to keep the integrity of the involved quantum states, we should have generation times much shorter than typical 
decoherence times, or $\gamma_{cav} \ll \Omega_m$ and $\gamma_{cav} \ll \eta G$. Although those are experimentally challenging conditions, 
we are witnessing recent significant advances in both ionic and optomechanical systems, even approaching the single-photon strong coupling 
regime in optomechanical systems \cite{marq14,aspelmeyer09,milburn10}. 

Finally, an important point is related to the verification of the existence of the generated entangled state. A direct measurement of
the entangled state may not be an easy task in general. As an alternative, we may envisage an entanglement witness, which could be 
implemented via local measurements. After accomplishing the generation of the entangled state in Eq. (\ref{entangstred}), for instance, 
we could carry out the following procedure: i) perform a projective measurement \cite{kim16} corresponding to 
$\hat{\Pi} = |1\rangle_v{}_v\langle 1|\otimes\hat{\mbox{I}}_m$ on the ionic vibrational state, thus collapsing the mirror's state to 
$|-\alpha_0\rangle_m$; ii) displace the mirror (via a classical light pulse) by an amplitude $\alpha_0$, which would bring the mirror 
to its ground (vacuum) state; iii) measure the mirror state \cite{aspelmeyer11} in order to check whether it is (or not) in its vacuum 
state. This would allows us to discriminate the entangled state in Eq. (\ref{entangstred}) 
from the pure product state 
$|\Phi\rangle = (|-\alpha_0 + 2\kappa\rangle_m + |-\alpha_0\rangle_m) \otimes (|0\rangle_v + |1\rangle_v)$, for instance.

It should be emphasized that entanglement herein involves three important features of the resulting entangled 
states: $(i)$ states are related to massive, although different physical systems, viz., a trapped ion and a vibrating mirror; 
$(ii)$ the entangled states are constituted by discrete states (ionic vibration) and continuous variable states (moving mirror), 
i.e., it is accomplished hybrid entanglement in spatially separated cavities; $(iii)$ states involve a microscopic system (ion) and 
a macroscopic system (mirror), also referred as a micro-macro entangled state. 

\begin{figure}
\centering
\resizebox{0.5\textwidth}{!}{
\includegraphics{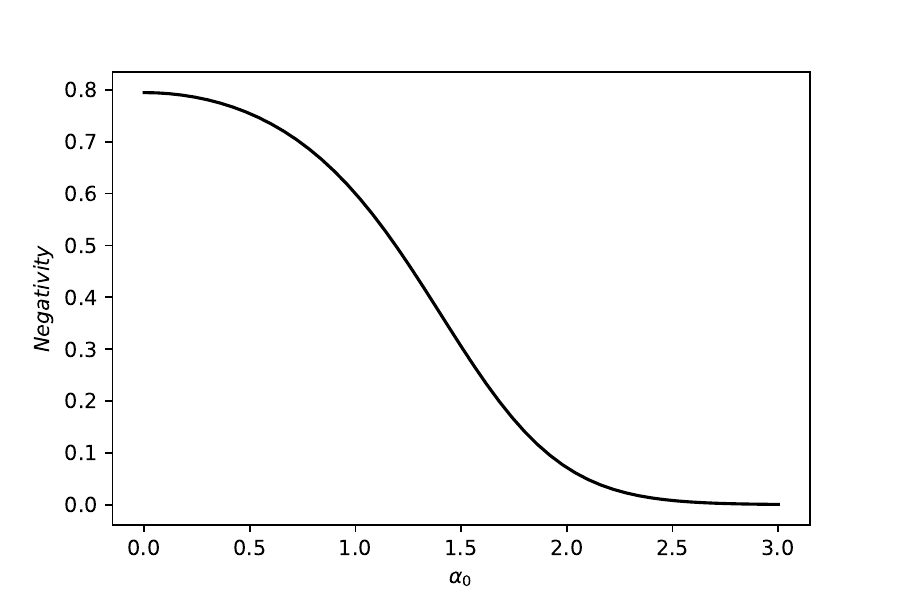}
}
\caption{Negativity relative to state in Eq. (\ref{entangstred}) as a function of $\alpha_0$ calculated at $t = 1.0\mu$s. 
The parameters of the optomechanical system have been taken as: $g = 5$ MHz, $\Omega_m = 10$ MHz, $\kappa = 0.5$, 
$\Gamma = 0.5$ MHz, while for the ionic system we have $\eta = 0.1$, $G = 5.0$ MHz, $\gamma = 0.025$ MHz, and 
$\theta = \pi/4$ rad.}
\end{figure}

\section*{Acknowledgements}

We would like to thank CNPq (Conselho Nacional para o 
Desenvolvimento Cient\'\i fico e Tecnol\'ogico) and FAPESP
(Funda\c c\~ao de Amparo \`a Pesquisa do Estado de S\~ao Paulo), Brazil,
for financial support through the National Institute for Science and 
Technology of Quantum Information (INCT-IQ under grant 2008/57856-6) and the 
Optics and Photonics Research Center (CePOF under grant 2005/51689-2).

\appendix

\section{Generation of ion-field states}

For the first red sideband case, the quantum jump evolution will be  
\[
	|\psi_{ion}^{r}(t')\rangle=\,{\cal{U}}_{ion}^{r}(t')|\psi_{ion}(0)\rangle
\]
which results in
\begin{eqnarray}
	\lefteqn{|\psi_{ion}^{r}(t')\rangle= \sin\theta|0\rangle_{1}|0\rangle_{v}|g\rangle }\nonumber \\
	&&+\cos\left(\eta Gt'\right)\cosh\left(\frac{\gamma}{2}\eta Gt'^{2}\right)\cos\theta|0\rangle_{1}|0\rangle_{v}|e\rangle \nonumber \\
	&&+i\cos\left(\eta Gt'\right){e}^{-\gamma t'}\sinh\left(\frac{\gamma}{2}\eta Gt'^{2}\right)\cos\theta|1\rangle_{1}|1\rangle_{v}|g\rangle  \nonumber \\
	&&+\sin\left(\eta Gt'\right){e}^{-\gamma t'}\sinh\left(\frac{\gamma}{2}\eta Gt'^{2}\right)\cos\theta|0\rangle_{1}|0\rangle_{v}|e\rangle  \nonumber \\	
	&&-i\sin\left(\eta Gt'\right)\cosh\left(\frac{\gamma}{2}\eta Gt'^{2}\right)\cos\theta|1\rangle_{1}|1\rangle_{v}|g\rangle.\nonumber
\end{eqnarray}

Now if the ion's electronic state is projected (via fluorescence, formally by applying $\hat{\Pi}_{g}=|g\rangle\langle g|$), we obtain
\[
	|\psi_{ion,g}^{r}(t')\rangle=\hat{\Pi}_{g}|\psi_{ion}^{r}(t')\rangle,
\]
\begin{eqnarray}
	\lefteqn{|\psi_{ion,g}^{r}(t')\rangle= N^{(r)}\Bigl[\sin\theta|0\rangle_{1}|0\rangle_{v}} \nonumber \\
	&&+i\cos\left(\eta Gt'\right){e}^{-\gamma t'}\sinh\left(\frac{\gamma}{2}\eta Gt'^{2}\right)\cos\theta|1\rangle_{1}|1\rangle_{v}  \nonumber \\
	&&-i\sin\left(\eta Gt'\right)\cosh\left(\frac{\gamma}{2}\eta Gt'^{2}\right)\cos\theta|1\rangle_{1}|1\rangle_{v}\Bigr].\nonumber
\end{eqnarray}

For $t'=\pi/2\eta G$, we have
\[
|\psi_{ion,g}^{r}(t'=\pi/2\eta G)\rangle= 
N^{(r)} \Bigl[\sin\theta|0\rangle_{1}|0\rangle_{v}-i\cosh\left(\frac{\pi^{2}\gamma}{8\eta G}\right)\cos\theta|1\rangle_{1}|1\rangle_{v}\Bigr].
\]

Analogously, in the first blue sideband case the quantum jump evolution will lead to 

\begin{eqnarray}
\lefteqn{|\psi_{ion}^{b}(t')\rangle= \cos\left(\eta Gt'\right)\cosh\left(\frac{\gamma}{2}\eta Gt'^{2}\right)\cos\theta
|0\rangle_{1}|1\rangle_{v}|e\rangle} \nonumber \\
&&+{e}^{i\xi}\sin\theta|0\rangle_{1}|1\rangle_{v}|g\rangle   \nonumber \\
&&-i\cos\left(\eta Gt'\right)e^{-\frac{\gamma t'}{2}}\sinh\left(\frac{\gamma}{2}\eta Gt'^{2}\right)\cos\theta|1\rangle_{1}|0\rangle_{v}|g\rangle  \nonumber \\
&&-\sin\left(\eta Gt'\right)e^{-\frac{\gamma t'}{2}}\sinh\left(\frac{\gamma}{2}\eta Gt'^{2}\right)\cos\theta|0\rangle_{1}|1\rangle_{v}|e\rangle  \nonumber \\	
&&-i\sin\left(\eta Gt'\right)\cosh\left(\frac{\gamma}{2}\eta Gt'^{2}\right)\cos\theta|1\rangle_{1}|0\rangle_{v}|g\rangle.
\end{eqnarray}
If the electronic state is projected on $|g\rangle$, and at $t' = \pi/2\eta G$ the resulting cavity-ion state will now read 

\begin{equation}
|\psi_{ion,g}^{b}(t'=\pi/2\eta G)\rangle=N^{(b)}\left[\sin\theta|0\rangle_{1}|1\rangle_{v}-ie^{-\frac{\gamma t}{2}} 
\cosh\left(\frac{\pi^{2}\gamma}{8\eta G}\right)\cos\theta|1\rangle_{1}|0\rangle_{v}\right].
\end{equation}

\end{document}